\documentclass[a4paper]{article}
\usepackage[psamsfonts]{amssymb}
\usepackage{amsmath}
\usepackage{epsfig}
\title{The Holographic Model of Dark Energy and Thermodynamics of Non-Flat Accelerated Expanding
Universe \vspace{1cm}}

\author{M. R. Setare \footnote{ e-mail: rezakord@ipm.ir}\\ S.Shafei\\ \\{\small Department of Physics,
University of Kurdistan, Sanandaj, Iran}}
\date{}
\begin{document}
\maketitle
\newcommand{\be}{\begin{equation}}
\newcommand{\ee}{\end{equation}}
\newcommand{\bq}{\begin{eqnarray}}
\newcommand{\eq}{\end{eqnarray}}
\vspace{1cm}

\abstract
 Motivated by recent results on non-vanishing spatial curvature \cite{curve} we employ
 the holographic model of dark energy to investigate the validity of first and
 second laws of thermodynamics in non-flat (closed) universe enclosed by
 apparent horizon $R_A$ and the event horizon measured from the
 sphere of horizon named $L$. We show that for the apparent
 horizon the first law is roughly respected for different epochs while the second laws of thermodynamics is respected
while for $L$ as the system's IR cut-off first law is broken down
and second law is respected for special range of  deceleration
parameter. It is also shown that at late-time universe $L$ is
equal to $R_A$ and the thermodynamic laws are hold, when the
universe has non-vanishing curvature. Defining the fluid
temperature to be proportional to horizon temperature the range
for coefficient of proportionality is obtained provided that the
generalized second law of thermodynamics is hold.

\newpage
\section{Introduction}
The accelerated expansion that based on recent astrophysical data
\cite{exp}, our universe is experiencing  is today's most
important problem of cosmology. Missing energy density - with
negative pressure - responsible for this expansion has been dubbed
Dark Energy (DE). Wide range of scenarios have been proposed to
explain this acceleration while most of them can not explain all
the features of universe or they have so many parameters that
makes them difficult to fit. The models which have been discussed
widely in literature are those which consider vacuum energy
(cosmological constant) \cite{cosmo} as DE, introduce fifth
elements and dub it quintessence \cite{quint} or scenarios named
phantom \cite{phant} with $w<-1$ , where $w$ is parameter of
state.

An approach to the problem of DE arises from holographic Principle
that states that the number of degrees of freedom related directly
to entropy scales with the enclosing area of the system. It was
shown by 'tHooft and Susskind \cite{hologram} that effective local
quantum field theories greatly overcount degrees of freedom
because the entropy scales extensively for an effective quantum
field theory in a box of size $L$ with UV cut-off $ \Lambda$. As
pointed out by \cite{myung}, attempting to solve this problem,
Cohen {\it et al.} showed \cite{cohen} that in quantum field
theory, short distance cut-off $\Lambda$ is related to long
distance cut-off $L$ due to the limit set by forming a black hole.
In other words the total energy of the system with size $L$ should
not exceed the mass of the same size black hole i.e. $L^3
\rho_{\Lambda}\le LM_p^2$ where $\rho_{\Lambda}$ is the quantum
zero-point energy density caused by UV cutoff $\Lambda$ and $M_P$
denotes Planck mass ( $M_p^2=1/{8\pi G})$. The largest $L$ is
required to saturate this inequality. Then its holographic energy
density is given by $\rho_{\Lambda}= 3c^2M_p^2/8\pi L^2$ in which
$c$ is free dimensionless parameter and coefficient 3 is for
convenience.

 As an application of Holographic principle in cosmology,
 it was studied by \cite{KSM} that consequence of excluding those degrees of freedom of the system
 which will never be observed by that effective field
 theory gives rise to IR cut-off $L$ at the
 future event horizon. Thus in a universe dominated by DE, the
 future event horizon will tend to constant of the order $H^{-1}_0$, i.e. the present
 Hubble radius. The consequences of such a cut-off could be
 visible at the largest observable scales and particulary in the
 low CMB multipoles where we deal with discrete wave numbers. Considering the power spectrum in finite
 universe as a consequence of holographic constraint, with different boundary
 conditions, and fitting it with LSS, CMB and supernova data, a cosmic duality between dark energy equation of state
 and power spectrum is obtained that can describe the low $l$ features extremely
 well.

 Based on cosmological state of holographic principle, proposed by Fischler and
Susskind \cite{fischler}, the Holographic Model of Dark Energy
(HDE) has been proposed and studied widely in the
 literature \cite{miao,HDE}. In \cite{HG} using the type Ia
 supernova data, the model of HDE is constrained once
 when c is unity and another time when c is taken as free
 parameter. It is concluded that the HDE is consistent with recent observations, but future observations are needed to
 constrain this model more precisely. In another paper \cite{HL},
 the anthropic principle for HDE is discussed. It is found that,
 provided that the amplitude of fluctuation are variable the
 anthropic consideration favors the HDE over the cosmological
 constant.

 In HDE, in order to determine the proper and well-behaved system's IR cut-off, there are some
difficulties that must be studied carefully to get results adapted
with experiments that claim our universe has accelerated
expansion. For instance, in the model proposed by \cite{miao}, it
is discussed that considering particle horizon, $R_p$,
 \be
 R_p=a\int_0^t\frac{dt}{a}=a\int_0^a\frac{da}{Ha^2}
\ee
 as the IR cut-off, the HDE density reads to be
 \be
  \rho_{\Lambda}\varpropto a^{-2(1+\frac{1}{c})},
\ee
 that implies $w>-1/3$ which does not lead to accelerated
universe. Also it is shown in \cite{easther} that for the case of
closed
universe, it violates the holographic bound.\\

The problem of taking apparent horizon (Hubble horizon) - the
outermost surface defined by the null rays which instantaneously
are not expanding, $R_A=1/H$ - as the IR cut-off in the flat
universe, was discussed by Hsu \cite{Hsu}. According to Hsu's
argument, employing Friedman equation $\rho=3M^2_PH^2$ where
$\rho$ is the total energy density and taking $L=H^{-1}$ we will
find $\rho_m=3(1-c^2)M^2_PH^2$. Thus either $\rho_m$ and
$\rho_{\Lambda}$ behave as $H^2$. So the DE results pressureless,
since $\rho_{\Lambda}$ scales as like as matter energy density
$\rho_m$ with the scale factor $a$ as $a^{-3}$. Also, taking
apparent horizon as the IR cut-off may result the constant
parameter of state $w$, which is in contradiction with recent
observations implying variable $w$ \cite{varw}. In our
consideration for non-flat universe, because of the small value of
$\Omega_k$ we can consider our model as a system which departs
slightly from flat space. Consequently we respect the results of
flat universe so that we treat apparent horizon only as an
arbitrary distance and not as the system's IR cut-off.

 On the other hand taking the event horizon, $R_h$, where
 \be
  R_h= a\int_t^\infty \frac{dt}{a}=a\int_a^\infty\frac{da}{Ha^2}
 \ee
 to be the IR cut-off, gives the results compatible with observations for flat
 universe.

 It is fair to claim that simplicity and reasonability of HDE provides
 more reliable frame to investigate the problem of DE rather than other models
proposed in the literature\cite{cosmo,quint,phant}. For instance
the coincidence or "why now" problem is easily solved in some
models of HDE based on this fundamental assumption that matter and
holographic dark energy do not conserve separately, but the matter
energy density
decays into the holographic energy density \cite{interac}.\\

Since the discovery of black hole thermodynamics in 1970
physicists have speculated thermodynamics of cosmological models
in accelerated expanding universe \cite{thermo}. Related to
present work, in \cite{abdalla}, for either time independent and
time-dependent equation of state (EoS), the first and second laws
of thermodynamics in flat universe were investigated. For the case
of constant EoS, the first law is valid for apparent horizon
(Hubble horizon) and it does not hold for event horizon as
system's IR cut-off. When the EoS is assumed to be time dependent,
using holographic model of dark energy in flat space, the same
result is gained: The event horizon, in contradict with apparent
horizon, does not satisfy the first law. Also, while the event
horizon does not respect the second law, it is hold for the
universe enclosed by apparent horizon.

Some experimental data has implied that our universe is not a
perfectly flat universe and recent papers have favored the
universe with spatial curvature \cite{curve}. As a matter of fact,
we want to remark that although it is believed that our universe
is flat, a contribution to the Friedmann equation from spatial
curvature is still possible if the number of e-foldings is not
very large \cite{miao2}.  Therefore, it would be interesting to
investigate the laws of thermodynamics for non-flat universe, and
determine for what distances, the thermodynamic laws, are
satisfied when the curvature is non-vanishing.

Defining the appropriate distance, for the case of non-flat
universe has another story. Some aspects of the problem has been
discussed in \cite{miao2,guberina}. In this case, the event
horizon can not be considered as the system's IR cut-off, because
for instance, when the dark energy is dominated and $c=1$, where
$c$ is a positive constant, $\Omega_\Lambda=1+ \Omega_k$, we find
$\dot R_h<0$, while we know that in this situation we must be in
de Sitter space with constant EoS. To solve this problem, another
distance is considered- radial size of the event horizon measured
on the sphere of the horizon, denoted by $L$- and the evolution of
holographic model of dark energy in non-flat universe is
investigated.

In present paper, using the holographic model of dark energy in
non-flat universe, we study the validity of first and second law
of thermodynamics in present time for a universe enveloped by
$R_A$ and $L$. In section 2, as the thermodynamic laws are
applicable in equilibrium,  we first investigate whether these
distances change dominantly over one hubble time, $t_H=1/H$, and
then we study the validity of first law. In section 3, the second
law of thermodynamics is studied. In final section, some
conclusions are represented.

We take $\hbar=k_B=G=c=1$.

\section{ First Law of Thermodynamics }
We consider the non-flat Friedmann-Robertson-Walker universe with
line element
 \be\label{metr}
ds^{2}=-dt^{2}+a^{2}(t)(\frac{dr^2}{1-kr^2}+r^2d\Omega^{2}).
 \ee
where $k$ denotes the curvature of space k=0,1,-1 for flat, closed
and open universe respectively. In non-flat universe, our choice
for holographic dark energy density is
 \be
  \rho_\Lambda=3c^2L^{-2}.
 \ee
As it was mentioned, $c$ is a positive constant in holographic
model of dark energy($c\geq1$)and the coefficient 3 is for
convenient. $L$ is defined as the following form:
\begin{equation}
 L=ar(t),
\end{equation}
here, $a$, is scale factor and $r(t)$ can be obtained from the
following equation
\begin{equation}\label{rdef}
\int_0^{r(t)}\frac{dr}{\sqrt{1-kr^2}}=\int_t^\infty
\frac{dt}{a}=\frac{R_h}{a},
\end{equation}
where $R_h$ is event horizon. For closed universe we have (same
calculation is valid for open universe by transformation)
 \be
 r(t)=\frac{1}{\sqrt{k}} sin y.
 \ee
where $y\equiv \sqrt{k}R_h/a$.
 The EoS of DE reads to be(relation obtained in \cite{miao2}):
\begin{equation}\label{eos}
w_\Lambda=\frac{-1}{3}(1+\frac{2}{c}\sqrt \Omega_{\Lambda}cosy).
\end{equation}
Here $\Omega_\Lambda$ is dimensionless DE density,
$\Omega_\Lambda=\rho_\Lambda / \rho_{cr}$. Putting its present
value $\Omega_{\Lambda}=0.73$, a lower bound for $w_\Lambda $, is
obtained to be $-0.90$, provided that $ c=1 $. If $c\geq 1$, then
$w_\Lambda$ will be always larger than $-1$. For $ c<1 $ ,
$\omega_{\Lambda} < -1$ and the holographic DE will have
phantom-like behavior, but imposing the Gibbons-Hawking entropy in
a closed universe results $c>1$ to avoid decreasing entropy. Thus
the holographic model of DE can not behave like phantom. For more
general bound on parameter $c$ see \cite{gong}. Our study is due
to present time, so the ordinary matter is taken into account. The
critical energy density, $\rho_{cr}$, DE density,
$\rho_{\Lambda}$, and the energy density of curvature, $\rho_k$,
are given by following relations respectively:
\begin{eqnarray} \label{ro}
\rho_{cr}=3H^2,\quad  \rho_{\Lambda}=3c^2L^{-2},\quad
\rho_k=\frac{3 k}{a^2}.
\end{eqnarray}
Using definition $\Omega_\Lambda$ and relation (\ref{ro}), $\dot
L$ gets: \be \label{ldot}
 \dot L= HL+ a \dot{r(t)}=\frac{c}{\sqrt{\Omega_\Lambda}}-cos y,
\end{equation}
so one can easily find that
\begin{equation}
t_H\frac{\dot L}{L}=1-\frac{\sqrt{\Omega_\Lambda}}{c} cos y.
\end{equation}
Clearly, $L$ does not change dominantly over one Hubble time, thus
the laws of thermodynamics can be applied here.

From \cite{bousso} the amount of energy crossing the surface
specified by $L$ during time interval $dt$ is obtained as
following
\begin{equation}
-dE=4\pi L^2 T_{ab}k^a k^b dt=4\pi L^2 \rho_{cr}(1+w)dt.
\end{equation}
where $k_a$ and $k_b$ are ingoing null vector fields and $w$ is
related to total pressure and energy density of matter enveloped
by horizon. We find
 \be\label{dEL}
 -dE=\frac{c^2}{\Omega_\Lambda}(\frac{3}{2}-\frac{\Omega_\Lambda}{2}-\frac{\Omega_\Lambda^{3/2}}{c}cos
 y),
 \ee
 to obtain this relation we have replaced $w$ by $w_\Lambda \Omega_\Lambda$.
Employing black hole thermodynamics, based on Bekenstein
\cite{bek}, Hawking and Gibbons works \cite{gib}, and generalizing
it to our cosmological horizons, we define the temperature and
entropy to be (for $L$):
  \be\label{TS}
 T_L=\frac{1}{2\pi L},\qquad S_L=\pi L^2
\ee Hence \bq \label{TLdS}
 T_LdS=\dot Ldt=(\frac{c}{\sqrt{\Omega_\Lambda}}-cos
 y)dt
\eq
 Comparing relations (\ref{dEL}) and (\ref{TLdS}) we see that
the first law of thermodynamics is not satisfied for $L$
\begin{equation}
-dE\neq TdS.
\end{equation}

On the other hand, we want to assert that redefining the
temperature of IR cut-off $L$ to be e.g. de Sitter temperature
 $T=H/2\pi$, will not change the invalidity of first law for the
 case of $L=ar(t)$ ( The calculation is straightforward as what has been performed in
 \cite{abdalla} for the case of flat universe.). So in continue our choice for the IR cut-off's temperature is
 $T_L=1/2\pi L$.\\

For the case of apparent horizon (Hubble horizon) - which we
consider it as an alternative distance to study our physical laws
in the universe enveloped by - we first investigate the ability of
applying thermodynamics laws. Using the following relation we find
an expression for $H^{-1}$
\begin{equation}\label{om}
\Omega_m=1+\Omega_k-\Omega_{\Lambda}=\frac{\rho_m}{3
H^2}=\Omega_m^0 H_0^2 H^{-2} a^{-3}
\end{equation}
where, $\rho_m= \Omega_m^0 \rho_{cr}^0 a^{-3}$ and $H_0$ denotes
the present value of Hubble parameter and
\begin{equation}\label{ok}
\Omega_k=\frac{\rho_k}{\rho_{cr}}=\frac{\Omega_k^0 \rho_{cr }^0
a^{-2}}{3H^2}=\Omega_k^0 H_0^2 H^{-2} a^{-2}.
\end{equation}
Using (\ref{om}) and (\ref{ok}) we find
\begin{equation}\label{H}
\frac{1}{H}=\frac{a^{3/2}}{\sqrt{\Omega_m^0}H_0}(\frac{1-{\Omega_\Lambda}}{1-a\gamma})^{1/2},
\end{equation}
here $\gamma \equiv{\Omega_k^0 / {\Omega_m^0}} <1$. Taking
derivative in both sides of (\ref{H}) with respect to $x(\equiv
lna)$, and after some calculation, we obtain
 \be \label{Hderiv}
H\frac{d}{dx}H^{-1}=1- \frac{\Omega_\Lambda^{3/2} cos y}{c}+
\frac{1-\Omega_{\Lambda}}{2(1-a\gamma)},
 \ee
where we have used following relation in which prime sign denotes
derivative respect to $x$ \cite{miao2}:
\begin{equation}
\Omega_{\Lambda}^\prime=\Omega_{\Lambda}(1-\Omega_{\Lambda})(\frac{2}{c}{\sqrt{\Omega_{\Lambda}}}cosy+
\frac{1}{1-a\gamma}).
\end{equation}
For $R_A$ we have
 \be
 t_h\frac{\dot R_A}{R_A}=H\frac{dH^{-1}}{dx}.
 \ee

 Fortunately from (\ref{Hderiv}) it is easily seen that we
are allowed to use thermodynamic laws for $R_A$. In fact neither
$R_A$ nor $L$ change dominantly over one Hubble time. Clearly, in
(\ref{Hderiv}), if $k\rightarrow 0$, then the result will be equal
to what has been obtained for the case of $R_A$ in flat universe.

Modifying relations (\ref{dEL}) and (\ref{TS}), for $R_A$, so that
 \bq
 T_A=\frac{1}{2\pi R_A}, \qquad S_A=\pi R^2_A. \nonumber
 \eq
We obtain the following relations:
 \be\label{dEA}
 -dE=(\frac{3}{2}-\frac{\Omega_\Lambda}{2}-\frac{\Omega^{3/2}_{\Lambda}}{c}cos
 y)dt
 \ee
 and
 \be\label{TAdSA}
  T_A dS_A=(1+\frac{1-\Omega_\Lambda}{2(1-a\gamma)}-\frac{\Omega^{3/2}_{\Lambda}}{c}cos
 y)dt.
 \ee

 At the first sight, it looks that the first law is violated.
 To  consider different situations precisely, we concentrate on the second term in parentheses of RHS
of relation (\ref{TAdSA}), where difficulties arise. Using some
approximation for this term, and applying that in the relation (8)
and comparing with the relation (\ref{dEA}), the validity of the
first law in different epochs can be studied. Roughly speaking, in
the early universe, where $a$ approaches to zero, we can write
 \be
\frac{1-\Omega_\Lambda}{2(1-a\gamma)}\simeq
\frac{1-\Omega_\Lambda}{2} \ee
 then first law is hold. At present-time we have, (taking $a=1$)
\begin{equation}\nonumber
 \frac{1-\Omega_\Lambda}{2(1-a\gamma)}= \frac{1-\Omega_\Lambda}{2(1-\gamma)},
\end{equation}
but noting that $\Omega_k^0=0.01$ and
$\Omega_m^0=1+\Omega_k^0-\Omega_\Lambda^0\simeq 0.28$ then $\gamma
\equiv{\Omega_k ^0/ {\Omega_m^0}}\thicksim 0.04$, one can ignore
$\gamma$ in denominator and
 \be
\frac{1-\Omega_\Lambda}{2(1-a\gamma)}\simeq
\frac{1-\Omega_\Lambda}{2},
\ee
 thereby the first law is roughly hold.
 Since for close future we can assume that $a\gamma$ is much
smaller than 1, then \be
\frac{1-\Omega_\Lambda}{2(1-a\gamma)}=\frac{1-\Omega_\Lambda}{2}(1+a
\gamma) , \ee and likewise the present time the first law is
approximately respected. Eventually at late-time universe where
$\Omega_\Lambda\simeq1$ the first law is hold provided that $a$
remains finite. Strictly speaking, only in late-time universe,
one can result that the first law of thermodynamics is hold with
no approximation.

As we saw, for $L$, as an the IR cut-off of our system, the first
law of thermodynamics did not hold in present time and non-flat
universe. Discussed  by \cite{abdalla}, the reason could be that
the first law is valid only when it is applied to nearby states of
local thermodynamics equilibrium and the IR cut-off that we
considered reflects global properties of the universe. Also, we
discussed above that the particle horizon did not work, due to it
does not lead to an accelerated universe and apparent horizon
(Hubble horizon) have some difficulties mentioned in introduction.
Therefore it looks that we need to define new distances or
redefine some of parameters- e.g Hawking temperature - so that the
thermodynamic laws are satisfied, although we remarked before,
that applying Hawking temperature $T=1/(2\pi R_A)$ for $L$ in
studying the validity of first law does not solve the problem, as
it did not solve the problem of event horizon, $R_h$, in flat
universe.

It is worthwhile to remind that one can investigate these
relations in open universe by transforming $k\rightarrow -k$,
$\rho_k \rightarrow -\rho_k$, $\Omega_k\rightarrow \Omega_k$,
$\gamma \rightarrow -\gamma$.

\section{Second Law of Thermodynamics}
Here, we study the validity of generalized second law (GSL) of
thermodynamics. According to GSL, for our system, the sum of the
entropy of matter enclosed by horizon and the entropy of horizon
must not be decreasing function of time. We investigate this law
for the universe filled with perfect fluid described by normal
scalar field (quintessence-like). For this purpose, we consider
the enclosed matter and calculate its entropy.

Before going into mathematics of GSL, we want to figure out
remarkable points about the temperature of the fluid. According to
generalization the black hole thermodynamics to our cosmological
model, we have taken the temperature of our horizon to be
$T_L=(1/2\pi L)$ where $L$ denotes the size of the universe. In
investigation the GSL, definition the temperature of the fluid
needs further discussions. The only temperature in hand is the
horizon temperature. If the fluid temperature is equal with the
horizon temperature, the system will be in equilibrium. Other
possibilities \cite{davies,mohseni} is that the fluid temperature
is proportional to horizon temperature i.e. for the fluid
enveloped by apparent horizon $T=bH/2\pi$. In continue, We shall
 show that if we want the generalized second law of thermodynamics to be hold,
redefining fluid temperature to be $T=bH/2\pi$ impose an upper
bound on $b$ to be $1$. But for now, it looks reasonable to take
$b=1$ \cite{pavon}.\\

The entropy of the matter has the following relation with its
pressure and energy
 \be
 dS=\frac{1}{T}\ (PdV+dE)
 \ee
 where $V$ is the volume containing the matter.

For the apparent horizon, we have $V=4\pi R_A^3/3$, $E=4\pi \rho
R_A^3/3=R_A/2$, $P=w\rho=w_\Lambda \Omega_\Lambda 3H^2/{8\pi}$,
then
 \be
 dS=\pi(1+3w_\Lambda \Omega_\Lambda)R_AdR_A.
\ee
 Using $R_AdR_A=-H^{-3}(dH/dx)dx$ one can obtain
\bq\label{dS1}
 \frac{d S}{d x}=\nonumber -\pi(1+3w_\Lambda
\Omega_\Lambda)H^{-3}\frac{dH}{dx}\\= \nonumber
-\pi(1-\Omega_\Lambda-\frac{2}{c}\Omega_\Lambda^{3/2} cos y
)H^{-3}\frac{dH}{dx}\\= -2\pi q H^{-3}\frac{dH}{dx}.
 \eq
Here $q$ is deceleration parameter defined as following
 \bq\label{q}
 q=-\frac{\dot H}{H^2}-1=-\frac{\Omega_\Lambda^{3/2} cos
 y}{c}+\frac{1-\Omega_\Lambda}{2(1-a\gamma)}\simeq
 \frac{1}{2}(1-\Omega_\Lambda-\frac{2}{c}\ \Omega_\Lambda^{3/2} cos y).
 \eq
The entropy of apparent horizon, is $S_A=\pi R_A^2$, so one can
easily find \be\label{dSA}
 \frac{dS_A}{dx}=-2\pi H^{-3}\frac{dH}{dx}.
\ee
 From the equations (\ref{dS1}) and (\ref{dSA}) it is obtained
 \be\label{dSTOTA}
 \frac{d}{dx}(S+S_A)=2\pi H^{-2}(\frac{dH}{dx})^2
 \ee
 which is clearly positive. Hence it is precisely concluded that GSL is
 respected for the sum of the entropy of apparent horizon and the
 entropy of matter enveloped by.

 As we mentioned before, we want to show that defining the fluid
 temperature
to be $T=bH/2\pi$, imposes upper bound on $b$ provided that the
GSL is satisfied. The equations will modify as follows:
 \bq
  dS=\frac{1}{T}\ (PdV+dE)=
  \frac{\pi}{b}(1+3w_\Lambda \Omega_\Lambda)R_AdR_A.
\eq
 One can find
\bq
 \frac{d S}{d t}=\frac{2\pi\dot H}{bH^{3}}(\frac{\dot H}{H^2}+1),
 \eq
 and
 \be
\frac{d S_A}{d t}=-\frac{2\pi\dot H}{H}
 \ee
 Thus it can be easily seen that
 \bq
 \frac{d}{dt}(S+S_A)=\frac{2\pi\dot H}{bH^3}(1-b+\frac{\dot
 H}{H^2}).
 \eq
If we note that for our model $\dot H<0$, then the term in
parentheses of RHS must be $\leq0$. Therefore a upper bound is
obtained for $b$ to be:
 \be
 b\geq 1+\frac{\dot H}{H^2}
 \ee
 As $\dot H<0$- assuming that we are in quintessence-like behavior- the
  RHS of inequality above is always smaller than or equal with
 1.\\

For $L$ we find the following relations
 \be \label{dSL}
 \frac{dS}{dx}=\frac{-\pi c^4}{H^2\Omega_\Lambda ^2}(\frac{1+3w_\Lambda
 \Omega_\Lambda}{H}\frac{dH}{dx}+\frac{3+3w_\Lambda
 \Omega_\Lambda}{2\Omega_\Lambda}\Omega'_\Lambda)
 \ee
 and
\be \label{dS2}
 \frac{dS_L}{dx}=-\frac{\pi
 c^2}{H^2\Omega_\Lambda^2}(2\frac{\Omega_\Lambda}{H}\frac{dH}{dx}+\Omega'_\Lambda).
 \ee
Using relations (\ref{dSL}) and (\ref{dS2}) we find
 \be\label{dSTOTL}
\frac{d}{dx}(S+S_L)=\frac{\pi c^4}{H^2\Omega_\Lambda
^2}\{(1+q)(2q+\frac{2\Omega_\Lambda}{c^2})+(\frac{1-\Omega_\Lambda}{\Omega_\Lambda})
(2q-1)(1+q+\frac{\Omega_\Lambda}{c^2})\}.
 \ee
We restrict our consideration to present time, $\Omega_\Lambda=
0.73$ and take $c$ to be 1 (changing $c$ slightly to get bigger
than 1, only moves slightly the range of amounts that $q$ can
take.) The sign of $\frac{d}{dx}(S+S_L)$ depends on the sign of
expressions $(2q+\frac{2\Omega_\Lambda}{c^2})$ and $(2q-1)$, since
other expressions are clearly positive. To determine the sign of
$q$, we pay attention that amounts that $q$ can take, depend on
the sign of $\dot H$. For $\dot H>0$ we have phantom-like behavior
and from previous discussions we know since the Hawking-Gibbon's
bound does not allow our holographic DE model to be of this kind,
we have to rule out this possibility. If $\dot H=0$ we are in de-
Sitter space-time and $q=-1$, therefore $\frac{d}{dx}(S+S_L)=0$.
Eventually for the case of $\dot H<0$ we find $q>-1$. Simplifying
(\ref{dSTOTL}) by means of putting values of $\Omega_\Lambda=0.73$
and $c=1$  makes the following form
 \be\label{ESTdS}
 \frac{d}{dx}(S+S_L)=2.74\ (0.22+q)(1.38+q).
 \ee
 From (\ref{ESTdS}) we find that for $q>-0.22$ then $\frac{d}{dx}(S+S_L)>0$  while
 for $q\in[-1,-0.22]$ we find $\frac{d}{dx}(S+S_L)<0$. Hence one can result, either in de
 Sitter space-time and for our accelerated model with $q> -0.22$ the
 generalized second law of thermodynamics is respected.

 In the end of this section we want to remark that at late-time
 universe, where $\Omega_\Lambda$ approaches to
 unity, provided that we take $c=1$, we find that $R_A=L$
 and using relations (\ref{dEL}, \ref{TLdS}, \ref{dEA}, \ref{TAdSA})
it is clear that the  the first law is satisfied. Also about the
second law, at late-time universe the relation (\ref{dSTOTL}) is
in absolute consistency with (\ref{dSTOTA}), therefore the second
law is also respected at late-time and non-flat universe.

\section{Summary}
In order to solve cosmological problems and because the lack of
our knowledge, for instance to determine what could be the best
candidate for DE to explain the accelerated expansion of universe,
the cosmologists try to approach to best results as precise as
they can by considering all the possibilities they have.
Investigating the principles of thermodynamics and specially the
second law- as global accepted principle in the universe - in
different models of DE, as one of these possibilities, has been
widely studied in the literature, since this investigation can
constrain some of parameters in studied models, say, P. C. Davies
\cite{davies} studied the change in event horizon area in
cosmological models that depart slightly from de Sitter space and
showed that for this models the GSL is respected for the normal
scalar field, provided the fluid to be viscous.

It is of interest to remark that in the literature, the different
scenarios of DE has never been studied via considering special
similar horizon, as in \cite{davies} the apparent horizon, $1/H$,
determines our universe while in \cite{mohseni} the universe is
enclosed by event horizon, $R_h$. As we discussed above for flat
universe the convenient horizon looks to be $R_h$ while in non
flat universe we define $L$ because of the problems that arise if
we consider $R_h$ or $R_p$ (these problems arise if we consider
them as the system's IR cut-off). Thus it looks that we need to
define a horizon that satisfies all of our accepted principles; in
\cite{odintsov} a linear combination of event and apparent
horizon, as IR cut-off has been considered.

In present paper, we studied $L$, as the horizon measured from the
sphere of the horizon as system's IR cut-off and apparent (or
Hubble) horizon. We investigated the first and second law of
thermodynamics at present time for the universe enveloped by this
horizons and obtained that for apparent horizon just like the flat
case, the first and second law in non-flat universe are respected
while for $L$ the first law did not hold and second law was
satisfied just for a range of $q$ which $q$ was deceleration
parameter. We related the invalidity  of the first law for $L$ due
to this point that $L$ reflects the global properties of the
system while the first law is related to local thermodynamics
equilibrium. Also, we showed that at late-time universe $L$ is
equal to $R_A$ and the thermodynamic laws are satisfied at this
time.

\section{Acknowledgment}
The authors would like to thank the referee because of his/her
useful comments, which assisted to prepare better frame for this
study.

\end{document}